\documentclass[usenatbib]{mn2e}
\usepackage{natbib}
\usepackage{apjfonts}
\usepackage{amsmath}

\usepackage{epsfig}
\usepackage{graphicx}
\usepackage{amssymb}
\usepackage{aas_macros}
\usepackage{color}

\newcommand{\kms}{\,{\rm km \, s^{-1}}}

\newcommand{\mpc}{\,{\rm Mpc}}
\newcommand{\pc}{\,{\rm pc}}
\newcommand{\oversim}[2]{\protect{\mbox{\lower0.5ex\vbox{%
   \baselineskip=0pt\lineskip=0.2ex
   \ialign{$\mathsurround=0pt #1\hfil##\hfil$\crcr#2\crcr\sim\crcr}}}}} 

\catcode`\"=\active\let"=\"  

\def\3{{\ss} }

\def\c12{{1\over 2}}

\def\d{{\rm d}}   
   
\def\plusplus{\raise 0.3ex\hbox{${\scriptstyle ++}$}{}}

\def\and{{{\rm M}31}}

\def\gyr{{\rm Gyr}}
\def\myr{{\rm Myr}}
\bibliography{biblio}

\begin{document}   
\title[Stellar envelopes of GCs with DM haloes]{Stellar envelopes of globular clusters embedded in dark mini-haloes}

\author[Jorge Pe\~{n}arrubia et al.]{Jorge Pe\~{n}arrubia$^{1}$\thanks{jorpega@roe.ac.uk}, Anna Lisa Varri$^{1}$, Philip G. Breen$^{2}$, Annette M. N. Ferguson$^{1}$, \newauthor Rub\'en S\'anchez-Janssen$^3$\\
  $^1$Institute for Astronomy, University of Edinburgh, Royal Observatory, Blackford Hill, Edinburgh EH9 3HJ, UK\\
  $^2$School of Mathematics and Maxwell Institute for Mathematical Sciences, University of Edinburgh, King's Buildings, Edinburgh EH9 3JZ, UK\\
 $^3$STFC UK Astronomy Technology Centre, Royal Observatory, Blackford Hill, Edinburgh, EH9 3HJ, UK
}
\maketitle  

\begin{abstract}
We show that hard encounters in the central regions of globular clusters embedded in dark matter (DM) haloes necessarily lead to the formation of gravitationally-bound stellar envelopes that extend far beyond the nominal tidal radius of the system.  
Using statistical arguments and numerical techniques we derive the equilibrium distribution function of stars ejected from the centre of a non-divergent spherical potential.
Independently of the velocity distribution with which stars are ejected, GC envelopes have density profiles that approach asymptotically $\rho\sim r^{-4}$ at large distances and become isothermal towards the centre.
Adding a DM halo component leaves two clear-cut observational signatures: (i) a flattening, or slightly increase of the projected velocity dispersion profile at large distances, and (ii) an outer surface density profile that is systematically shallower than in models with no dark matter.
\end{abstract}


\section{Introduction}\label{sec:intro}
The formation of globular clusters (GCs) in a cosmological context remains an open issue. Scenarios of GC formation can be divided in two broad categories: (i) {\it primeval models}, where GCs originate as gravitationally-bound gas clouds in the early Universe (Peebles \& Dickie 1968; Kravtsov \& Gnedin 2005; Kruijssen 2015), (ii) {\it galactic origin}, where GCs are formed in dark matter mini-haloes before, or shortly after re-ionization begins (e.g. Peebles 1984; Bromm \& Clarke 2002; Mashchenko \& Sills 2005; Ricotti et al. 2016). Bekki \& Yong (2012) propose an intermediate scenario, where GCs correspond to the remnants of tidally-stripped nucleated galaxies.
 
The detection of diffuse, spherical stellar envelopes that extend out hundreds of parsecs around GCs suggests that at least some GCs may be embedded in dark matter (DM) haloes (Olszewski et al. 2009; Kuzma et al. 2016). This scenario has gained further support from recent spectroscopic surveys that reveal 
the presence of ``extra-tidal'' or ``halo stars''  with kinematics and chemical compositions consistent with those exhibited by the parent cluster (Marino et al. 2014), but located many times beyond the nominal tidal radius of the clusters (Kunder et al. 2014; Navin et al. 2015, 2016). In addition, the outskirts of some clusters exhibit flattened velocity dispersion profiles (e.g. Lane et al. 2010), although this behaviour can also be explained by tidal heating (e.g.  Kundic \& Ostriker 1995), and/or a population of potential escapers (e.g. Daniel et al. 2017).

Testing the existence of DM in GCs is complicated by the presence of non-luminous baryonic matter, such as white dwarfs, neutron stars and black holes, which may comprise a significant fraction of the cluster mass (Heggie \& Hut 1996). Also, GCs orbiting in the inner regions of the Galaxy may lose a large fraction of the primordial DM halo to tides (Bromm \& Clarke 2002). 
Attempts to infer extended DM envelopes in remote GCs, such as NGC 2419\footnote{However, a relatively large Galactocentric distance may not be a sufficient condition to rule out tidal stripping. E.g. Palomar 14, at a similar distance, shows a morphology reminiscent of tidal tails (Sollima et al. 2011).}, show kinematics, surface brightness and mass-to-light ratios that do not indicate a significant amount of DM inside the nominal tidal limit (Baumgardt et al. 2009; Ibata et al. 2013).
An independent argument against the presence of DM in NGC 2419 was presented in Conroy et al. (2011), who model the effects of two-body (soft) encounters as a diffusion process in phase-space using the stochastic theory of Spitzer \& Shapiro (1972). In clusters with no DM the predicted profile 
approaches asymptotically $\rho\sim r^{-3.5}$ at large radii, becoming systematically shallower as the mass of the DM halo component increases. These models do not match the observed (de-projected) density profile, which roughly scales as $\rho\sim r^{-4}$ far from the cluster centre (Bellazzini 2007). A similar behaviour has been found in several GCs of the Milky Way (e.g. Carballo-Bello et al. 2012) and M31 (Mackey et al. 2010).

Here we use statistical (\S\ref{sec:stats}) and numerical (\S\ref{sec:nbody}) methods to study the equilibrium configuration of stars ejected from the central regions of GCs with DM mini-haloes as a result of {\it hard} encounters with binary stars and/or intermediate-mass black holes. This type of encounters can propel particles with speeds that often exceed the central escape velocity of the cluster and may explain the detection of high-velocity stars in the core of NGC 2808 (Luetzgendorf et al. 2012), M3 and M13 (Kamann et al. 2014). 
Adding a DM halo component increases the number of centrally-ejected stars that remain gravitationally bound to the system. In \S\ref{sec:results} we show that  these stars form an isotropic envelope that extends far beyond the stellar size of the cluster. The DM halo leaves observational signatures in the (projected) distribution and line-of-sight velocities of the cluster outskirts, which we discuss in \S\ref{sec:summary}.

\section{Statistical model}\label{sec:stats}
Let us write the distribution function for an ensemble of stars ejected from the cluster centre as at $t=0$ as 
\begin{eqnarray}\label{eq:f0}
f({\bf r},v_r,v_t,t=0)=\delta({\bf r})p(v_r)\delta(v_t),
\end{eqnarray}
where $p(v_r)$ is the probability that a star is ejected with a velocity $v_r$ in the interval $v_r,v_r+\d v_r$, and $\delta$ is the Dirac's delta function. Here we are mostly interested in stars bound to the cluster, hence we shall limit our analysis to $v_r\le v_e(0)=\sqrt{-2\Phi_0}$, where $\Phi_0\equiv \Phi({\bf r}=0)$, bearing in mind that particles with $v_r>v_e(0)$ will drift away from the cluster on hyperbolic orbits. Also, for simplicity we choose a minimum velocity $v_{r,{\rm min}}\approx 0$, such that the distributin $p(v_r)$ is non-zero within the range $v_r\in(0,v_e(0)]$.

Clearly, a stellar ensemble that follows Equation~(\ref{eq:f0}) is far from dynamical equilibrium. Indeed, initially all particles are centrally located and have radial velocities, $v_r\ge 0$, which translates into a positive flux of stars from the centre outwards. However, on time-scales $t\gg t_{\rm cross}=r_c^{3/2}/(GM_c)^{1/2}$, where $M_c$ and $r_c$ are the cluster mass and half-light radius, respectively, phase-space mixing will bring energetically-bound particles into dynamical balance, where the number of stars moving outwards equals that moving inwards. 

To calculate the {\it equilibrium} (i.e. phase-mixed) distribution function, $f_{\rm eq}({\bf r},{\bf v})\equiv \lim_{t\to\infty} f({\bf r},{\bf v},t)$, let us also assume that the cluster has a spherical shape and is in isolation, such that the energy ($E$) and the angular momentum (${\bf L}$) are conserved quantities. 
From Equation~(\ref{eq:f0}) the probability to find a particle with integrals of motion within the interval $(E,E+\d E)$ and $(L,L+\d L)$ is 
\begin{eqnarray}\label{eq:nel}
N(E,L) = \frac{p[v_r(E)]}{[2(E-\Phi_0)]^{1/2}}\delta(L),
\end{eqnarray}
where $E=v_r^2/2 + \Phi_0$.

For mixed particle ensembles the equilibrium distribution function is found by mapping points in the integral-of-motion space onto the phase-space volume $\d^6\Omega=\d^3{\bf r}\d^3{\bf v}$ and taking into account that $N(E,L$) is a dynamical invariant 
\begin{eqnarray}\label{eq:feq}
f_{\rm eq}({\bf r},{\bf v})\d^6 \Omega = f_{\rm eq}(E,L)\omega(E,L)\d E\d L=
N(E,L)\d E\d L,
\end{eqnarray}
where $\omega$ is the so-called {\it density of states} and defines the maximum phase-space volume that particles with a given combination of integrals of motion can potentially sample (see Appendix A of Pe\~narrubia 2015 for details). For spherical systems
\begin{eqnarray}\label{eq:omega}
\omega(E,L)=8 \pi^2 L P(E,L),
\end{eqnarray}
 where $P(E,L)=2\int_{r_p}^{r_a} \d r/v_r$ is the period of an orbit with peri- and apo-centres $r_p$ and $r_a$, respectively. Combination of Equations~(\ref{eq:nel})~(\ref{eq:feq}) and~(\ref{eq:omega}) yields
\begin{eqnarray}\label{eq:feq2}
f_{\rm eq}(E,L)= f_0\frac{p[v_r(E)]}{[2(E-\Phi_0)]^{1/2}}\frac{\delta(L)}{8\pi^2 LP(E,L)},
\end{eqnarray}
where $f_0$ is a normalization constant that guarantees $\int f_{\rm eq}\d^6 \Omega=1$.

The associated density profile can be straightforwardly calculated from Equation~(\ref{eq:feq2}) by writing the volume element in velocity space as $\d^3 v= 2\pi \d E L\d L/(r^2 v_r)$, where $v_r^2=2[E-\Phi(r)-L^2/(2r^2)]$ (Pe\~narrubia 2015), and marginalizing over $\d L$ , which yields
\begin{eqnarray}\label{eq:rho}
\rho(r)\equiv \int d^3v f_{\rm eq}= \frac{f_0}{4\pi r^2}\int_{\Phi(r)}^0\d E\frac{p[v_r(E)]}{[2(E-\Phi_0)]^{1/2}}\frac{1}{[2(E-\Phi)]^{1/2}P(E,0)}.
\end{eqnarray}

The radial velocity dispersion profile can be written as
\begin{eqnarray}\label{eq:sr}
\sigma_r^2(r)\equiv \frac{\int d^3v v_r^2 f_{\rm eq}}{\int d^3v f_{\rm eq}}= \frac{f_0}{4\pi r^2\rho(r)}\int_{\Phi(r)}^0\d E\frac{p[v_r(E)]}{[2(E-\Phi_0)]^{1/2}}\frac{[2(E-\Phi)]^{1/2}}{P(E,0)},
\end{eqnarray}
while $\sigma_t=0$ by construction.

The asympotic behaviour of the profiles at $r\ll r_c$ depends on whether or not the integrals in~(\ref{eq:rho}) and~(\ref{eq:sr}) converge in the limit $r\to 0$. E.g. if the limit $\Phi_0=\lim_{r\to 0}\Phi(r)$ exists, then $\int_{\Phi(r)}^0d E\to \int_{\Phi_0}^0\d E$ in~(\ref{eq:rho}) and~(\ref{eq:sr}), such that $\rho(r)\sim r^{-2}$ and $\sigma_r\sim {\rm const.}$ towards the centre of the potential. Hence, the equilibrium configuration becomes isothermal in the inner-most regions of the cluster.

At large distances, $r\gg r_c$, the cluster potential approaches $\Phi\simeq -GM_c/r$. The orbital period of radial orbits in a Keplerian potential can be expressed analytically as $P(E,0)=2\pi GM_c/(-2 E)^{3/2}$. Note that only stars that are loosly bound to the cluster, $E\approx 0$, probe such large distances. For these particles $E-\Phi_0\simeq -\Phi_0$, and the velocity distribution becomes constant, $p[v_r(E)]\simeq p(\sqrt{-2\Phi_0})$, which can be taken out of the integral~(\ref{eq:rho}), returning a density profile that scales as $\rho\sim r^{-4}$ at $r\gg r_c$. Following similar steps, we find from Equation~(\ref{eq:sr}) that the velocity dispersion approaches asymptotically the Keplerian profile $\sigma_r\sim r^{-1/2}$ far from the cluster centre. 

It is worth stressing that, insofar as $\Phi(r)$ does not diverge, the asymptotic limits for $\rho(r)$ and $\sigma_r(r)$ derived above hold independently of the velocity distribution with which stars are ejected from the cluster centre, i.e. the function $p(v_r)$ defined in Equation~(\ref{eq:f0}). Indeed, previous work shows that equilibrium spherical, self-gravitating systems with a finite total mass have profiles that approach $\rho\sim r^{-4}$ at large distances (Jaffe 1987; Makino et al. 1990; Aguilar 2008).
  Below we inspect this issue in more detail with the aid of test-particle experiments.
  

\begin{figure*}
\begin{center}
\includegraphics[width=165mm]{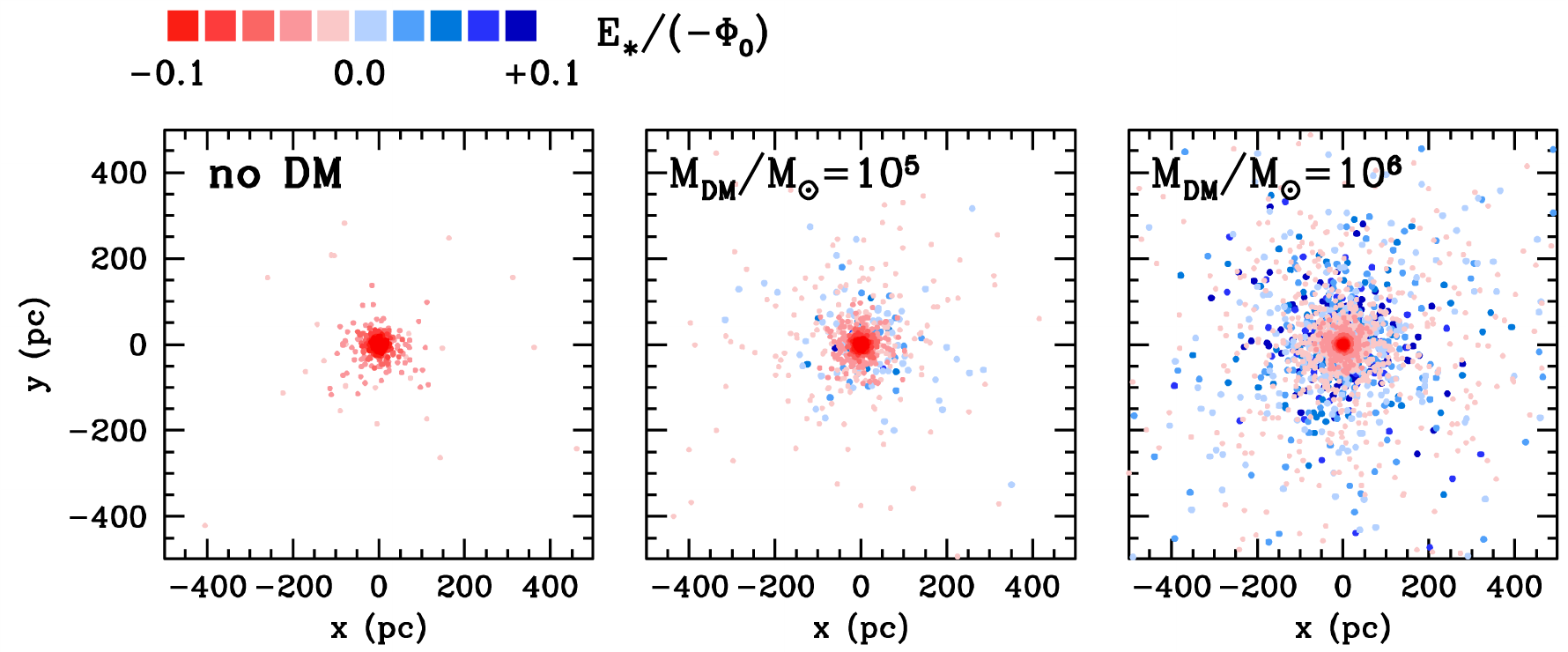}
\end{center}
\caption{Projected distribution of bound ($E<0$) stars ejected from the centre of a cluster with $M_\star=10^4M_\odot$ and different amounts of DM. In these models particles are ejected with a flat ($\alpha=0$) velocity distribution (see text). Particles are colour-coded according to the energy $E_\star=1/v^2+\Phi_\star$. Red (blue) colours denote stars bound to the stellar (stellar plus DM) potential. Note that stellar haloes can reach out to several hundred parsecs in GCs with DM haloes. }
\label{fig:xyz}
\end{figure*}
\section{Numerical experiments}\label{sec:nbody}
Following Leonard \& Tremaine (1990) let us adopt a power-law velocity distribution, $p(v_r)\propto v_r^{-\alpha}$, with $\alpha\ge 0$.
Using the distribution function~(\ref{eq:f0}) we generate samples of $N=2\times 10^5$ particles with velocities that fall in the interval $0< v_r\le v_e(0)$ and directions that are randomnly oriented on the surface of a sphere. 

We consider two-component cluster models where $M_c=M_\star+M_{\rm DM}$ and $\Phi=\Phi_\star+\Phi_{\rm DM}$. For simplicity, we adopt a Plummer (1912) model for the stellar component
\begin{eqnarray}\label{eq:phis}
\Phi_\star(r)=-\frac{G M_\star}{\sqrt{r^2+a^2}},
\end{eqnarray}
with a fixed mass $M_\star=10^4M_\odot$ and scale radius $a=2\pc$. The half-mass radius is $r_c\simeq 1.3a$. The DM halo is represented with a Hernquist (1990) potential
\begin{eqnarray}\label{eq:phidm}
\Phi_{\rm DM}(r)=-\frac{G M_{\rm DM}}{r+r_{\rm DM}},
\end{eqnarray}
which roughly matches the potential of CDM haloes orbiting in a more massive host (e.g. Pe\~narrubia et al. 2010). The size of the scale radius $r_{\rm DM}$ is chosen according to the mass-concentration relation found by Prada et al. (2012) at redshift $z=0$ in a flat Universe wtih $\Omega_m=0.7$ and $H_0=70\kms\mpc^{-1}$, which yields $r_{\rm DM}/\pc= 52.4, 132.5$ for $M_{\rm DM}/M_\odot=10^5$ and $10^6$, respectively.
We emphasize that these models are chosen for illustration, as the DM distribution in GCs remains unknown.

A few aspects of these systems are worth highlighting. Note first that the DM haloes are far more extended than the stellar component, $r_{\rm DM}/a\sim 25$--65, which suggests that centrally-ejected stars may distribute far beyond the nominal tidal radius of the cluster. Second, although we consider DM haloes with masses $M_{\rm DM}\gg M_\star$, the amount of DM in the central regions of the clusters is negligible, $2 M_{\rm DM}(<r_c)/M_\star\simeq 0.04$ and 0.07, for $M_{\rm DM}/M_\odot=10^5$ and $10^6$, respectively, where $M_{\rm DM}(<r)=M_{\rm DM}[r/(r+r_{\rm DM})]^2$. In contrast, the contribution of the DM halo to the central escape velocity is significant, $v_e(0)/v_{e,\star}(0)=[1+(M_{\rm DM}/M_{\star})(a/r_{\rm DM})]^{1/2}=1.17$ and 1.58 for $M_{\rm DM}/M_\odot=10^5$ and $10^6$, respectively, where $v_{e,\star}(0)=\sqrt{-2\Phi_\star(0)}=6.55\kms$.

The equations of motion $\ddot {\bf r}=-\nabla \Phi({\bf r})$ are solved for individual particles using a Runge-Kutta scheme (e.g. Press et al. 1992) with a variable time-step set such that energy conservation in isolation is better than $1:1000$. To guarantee dynamical equilibrium we integrate the particle ensemble for $10\gyr$, which is approximately 250 times longer than the crossing times of our dark matter halo models, $t_{\rm cross}=r_{\rm DM}^{3/2}/\sqrt{GM_{\rm DM}(<r_{\rm DM})}\sim 40 \myr$.

\section{Results}\label{sec:results}

Fig.~\ref{fig:xyz} shows 
the spatial distribution of centrally-ejected stars for cluster models with no DM (left-hand panel) and models embedded in DM haloes with masses $M_{\rm DM}=10^5 M_\odot$ (middle panel) and $M_{\rm DM}=10^6 M_\odot$ (right-hand panel). Particles are colour-coded according to the energy $E_\star=1/2v^2+\Phi_\star$, where red/blue denote negative/positive energies. As expected, equilibrium models with extended DM haloes contain gravitationally-bound particles located well beyond the visible size of the cluster. However, a large fraction of those particles are not bound to the stellar potential, i.e. $E_\star=v^2/+\Phi_\star>0$, and would be therefore tagged as ``extra-tidal'', or ``halo'' stars in models that neglect the presence of DM.
Comparison between models shows that the size of the stellar halo is directly correlated with the amount of DM in the cluster.

\begin{figure*}
\begin{center}
\includegraphics[width=145mm]{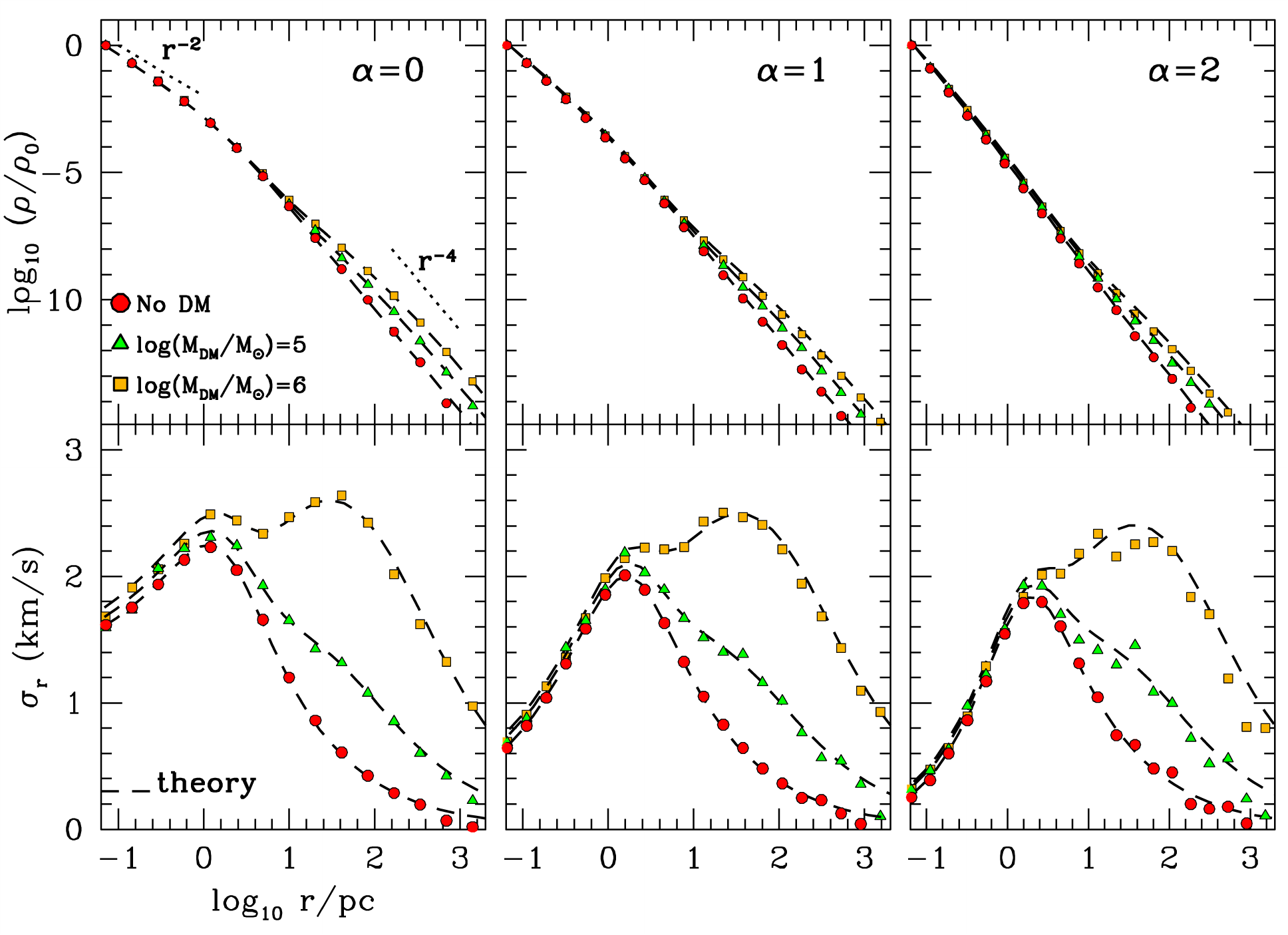}
\end{center}
\caption{Density profile of centrally-ejected stars in the cluster potential. Symbols denote test-particle models with different DM halo masses. Black-dashed lines show the density and velocity dispersion profiles given by Equations~(\ref{eq:rho}) and~(\ref{eq:sr}), respectively. Upper panels indicate that $\rho\sim r^{-4}$ at $r\gg a=2 \pc$ independently of the shape of the velocity distribution ($\alpha$). Note that the presence of a DM halo inflates the velocity dispersion profile at large distances.}
\label{fig:prof}
\end{figure*}

Fig.~\ref{fig:prof} highlights the excellent agreement between the spatial and kinematical distributions derived from the equilibrium distribution function $f_{\rm eq}(E,L)$ in Equation~(\ref{eq:feq2}) and those measured from the test-particle simulations. Upper and lower panels plot, respectively, the density and velocity dispersion profiles of particles ejected from the cluster centre with an initial velocity distribution $p(v_r)\sim v_r^{-\alpha}$, with coloured-symbols denoting test-particle results, and black-dashed lines the values returned by Equations~(\ref{eq:rho}) and~(\ref{eq:sr}). Each column corresponds to a different value for the power-law index, $\alpha$.
As predicted in Section~\ref{sec:stats}, centrally-ejected stars in dynamical equilibrium approach asymptotically $\rho\sim r^{-2}$ towards the centre of the potential, rolling towards $\rho\sim r^{-4}$ in the limit $r\to \infty$.
The presence of an extended stellar envelope can be seen in the upper panel of Fig.~\ref{fig:prof} as a systematic increase in the value of $\rho(r)$ at fixed radii $r\gtrsim r_c$ with respect to cluster models devoid of DM. As expected, incresing the power-law index decreases the probability to find particles ejected with high-velocities. This leads to steep outer profiles or, equivalently, centrally-concentrated stellar haloes. Indeed, the visible scatter in the binned profiles in the lower-right panel results from the decreasing number of particles located at large distances.

The velocity dispersion of the stellar haloes exhibit a remarkable sensitivity to the DM halo potential. In particular, models devoid of DM have velocity dispersion profiles that approach asymptotically the Keplerian curve $\sigma_r\sim r^{-1/2}$ at $r\gtrsim r_c$. The DM halo potential inflates the outer velocity dispersion profile, shifting the Keplerian behaviour to radii $r\gtrsim r_{\rm DM}\gg r_c$. Interestingly, the shape of $\sigma_r(r)$ at large radii is barely sensitive to the power-law index of the velocity distribution with which particles are ejected. This dependence is mostly visible at small radii, where $\sigma_r$ increases in models with $\alpha\sim 0$. At small radii the velocity dispersion profile becomes flat, as expected from the results of 
Section~\ref{sec:stats}. This behaviour is clearly visible in the lower-left panel of Fig.~\ref{fig:prof}, whereas the middle and right panels only show hints of a central flattening of $\sigma_r$ due to the limited resolution of our test-particle experiments.

\begin{figure}
\begin{center}
\includegraphics[width=84mm]{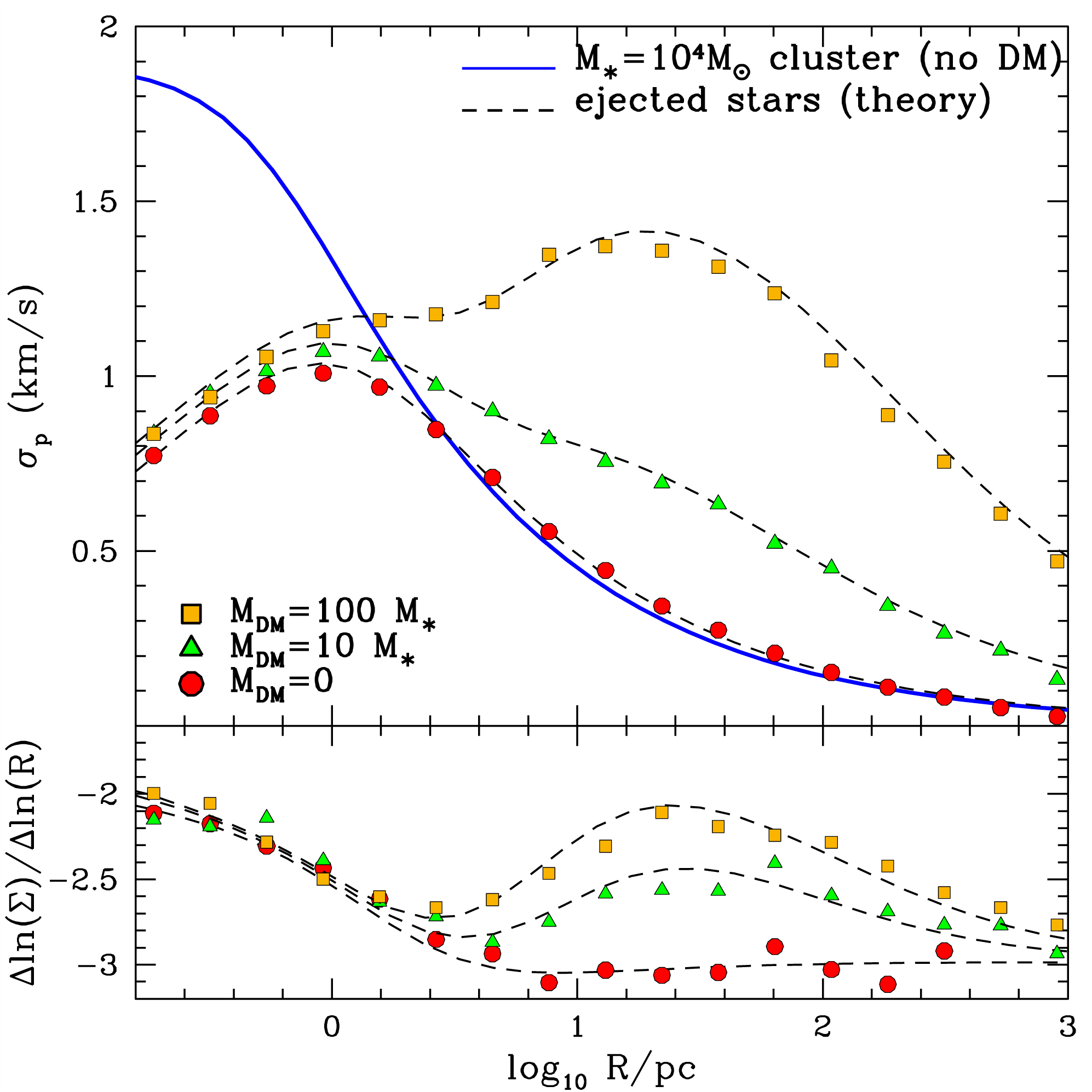}
\end{center}
\caption{{\it Upper panel:} Projected velocity dispersion profiles of stars ejected from the centre of cluster models with $\alpha=1$ for different DM halo masses. For ease of reference we plot the dispersion of an isotropic Plummer sphere with $M_\star=10^4 M_\odot$ and scale radius $a=2\pc$ (blue-solid line). Dashed solid lines show the velocities derived from the distribution function~(\ref{eq:feq2}). Note that the velocity dispersion of centrally-ejected stars follow a Keplerian decay at $\sigma_p(R)\sim R^{-1/2}$ at $R\gg a$. In contrast, the presence of a DM envelope systematically inflates the outer velocity dispersion profile. {\it Lower panel:} Power-law index of the surface density profile as a function of projected radius. Adding a DM component leads to outer profiles that are systematically shallower than in models with no DM.}
\label{fig:proj}
\end{figure}

Direct measurement of 6D phase-space coordinates is currently unfeasible in the majority of clusters owing to their large heliocentric distances. To investigate whether the projected spatial and kinematical distributions contain information on the amount of DM in GCs we plot in the upper panel of Fig.~\ref{fig:proj} the line-of-sight velocity dispersion $\sigma_p^2(R)=2\Sigma^{-1}(R)\int_R^\infty\d r \rho(r)\sigma_r^2(r)\sqrt{1-R^2/r^2}$. Here, $\Sigma(R)=2\int_R^\infty\d r r\rho(r)/\sqrt{r^2-R^2}$ is the surface density, while $\sigma_t=0$ by construction.
For the sake of clarity we show models with $\alpha=1$, noting that our conclusions do not depend on this particular choice. Note also that the outer dispersion profile of centrally-ejected stars in models devoid of DM (red symbols) is very similar to the velocity dispersion of a self-gravitating, isotropic Plummer sphere (blue-solid line). At small distances the shape of $\sigma_p(R)$ depends on the parameter $\alpha$. In general, we find that centrally-ejected stars tend to be colder than the self-gravitating stellar component.

The presence of an extended DM halo yields two clear-cut observational signatures. First, at large radii the velocity dispersion flattens out, or rises slightly, instead of decaying as $\sigma_p\sim R^{-1/2}$, as observed in models with no DM. E.g. at $R=100\pc$ we find that the velocity dispersion increases from $\sigma_p(100\pc)\simeq 0.16\kms$ in clusters with $M_{\rm DM}=0$, up to $0.46\kms$ and $1.14\kms$ for models with DM halo masses $M_{\rm DM}=10^5$ and $10^6M_\odot$, respectively. 

A second tell-tale of a DM halo envelope can be found in a relatively shallow power-law index of the outer profile, $\gamma=\Delta \ln \Sigma/\Delta \ln R$, shown in the lower panel of Fig.~\ref{fig:proj}. Recall that at very large radii the density profile of centrally-ejected stars scales as $\rho\sim r^{-4}$, hence we expect $\Sigma\sim R^{-3}$ at $R\gg r_c$. Indeed, we find that models with no DM (red symbols) have $\gamma\simeq -3$ at $R\gtrsim 10\pc$. In contrast, models with DM haloes exhibit considerably shallower slopes. E.g. at $R=100\pc$ the slope of cluster models with $M_{\rm DM}/M_\star=10$ and $100$ is $\gamma(100\pc)\simeq -2.6$ and $-2.3$, respectively.

\section{Discussion \& Summary}\label{sec:summary}
This paper inspects the equilibrium configuration of stars ejected from the central regions of a spherical cluster.
We find that in models with a non-divergent Newtonian potential the equilibrium profile approaches asymptotically $\rho\sim r^{-2}$ at $r\ll r_c$, and $\rho\sim r^{-4}$ at $r\gg r_c$, where $r_c$ is the cluster half-light radius, independently of the velocity distribution with which stars are ejected.


Adding a DM halo potential increases the maximum distances that energetically-bound ($E<0$) particles can reach, leading to an equilibrium configuration that extends far beyond the stellar size of the cluster.
A large fraction of these particles are not bound to the stellar component ($E_\star=1/2v^2+\Phi_\star=E-\Phi_{\rm DM}>0$), which motivates the label of ``extra-tidal'', or ``halo stars'' in the literature.

Despite its negligible contribution to the mass enclosed within the stellar radius of our GC models, the presence of a DM halo leaves two clear-cut observational signatures at large radii:
(i) a flattening, or slightly increase of the line-of-sight velocity dispersion profile $\sigma_p(R)$ at $R\gtrsim r_c$, and (ii) an outer surface density profile with a power-law index $\gamma=\Delta \ln \Sigma/\Delta \ln R>-3$ that is systematically shallower than in cluster models with no DM.


Unfortunately, it is difficult to predict the number of centrally-ejected stars that end up populating the halo, as this depends on quantities that remain largely unconstrained, such as the stellar mass function, the fraction of primordial binaries and the existence of intermediate-mass black holes. Given that the probability of hard encounters scales with the mean density of the system, we expect the majority of halo stars to be ejected during the virialization of proto-clusters and the subsequent core collapse(s). 
In addition, external processes may also impact on the halo luminosity.  
E.g. our scenario predicts the absence of stellar haloes in clusters with tidal streams, as these have lost their dark \& stellar envelopes to tides.

The detection \& characterization of stellar haloes is complicated by the low surface brightness of these structures. E.g. observations of M2 (Kuzma et al. 2016) and NGC 1851 (Olszewski et al. 2009) show spherical stellar envelopes that extend out for hundreds of parsces beyond the nominal tidal radii of the clusters, but only comprise $\sim 1\%$ of the luminosity of the entire system.
Deep, wide-field photometric data combined with future spectroscopic surveys (e.g. WEAVE, 4MOST) will enable a first assessment of the ubiquity of GC envelopes and help to constrain their origin. E.g. we expect centrally-ejected stars to have systematic lower masses than the overall cluster population (Vesperini \& Heggie 1997), but a similar chemical composition, which may be difficult to accommodate in models where GC envelopes correspond to the tidal remnants of nucleated dwarf galaxies (e.g. Norris et al. 2014).


The analytical distribution function derived in this paper may provide a useful tool to infer the DM content of GCs via Bayesian modelling the spatial and kinematic distribution of individual halo stars.
However, our analysis relies on conservation of energy and angular momentum, which may not be a realistic assumption in collisional systems owing to relaxation effects and/or the presence of an external tidal field. 
Studying these processes goes beyond the goals of this letter and calls for self-consistent test-particle simulations that follow the evolution of clusters embedded in live DM haloes orbiting around a massive host galaxy (Breen et al. {\it in prep.}).


\section{acknowledgements}
We are grateful to D. C. Heggie for crucial suggestions and to the referee L. Aguilar for an insightful report. J. Peacock and A. Meiksin are also warmly thanked for stimulating conversations. This work has benefited from discussions held at the Royal Society meeting ``Globular Cluster Formation and Evolution in the Context of Cosmological Galaxy Assembly'. ALV acknowledges support from the EU Horizon 2020 programme (MSCA-IF-EF-RI 658088) and PGB from the Leverhulme Trust (RPG-2015-408).

\end{document}